# Postsingular Science

Eldar Knar[1]

Institute of Philosophy, Political Science and Religion Studies
Ministry of Science and Higher Education of the Republic of Kazakhstan
eldarknar@gmail.com
https://orcid.org/0000-0002-7490-8375

**Abstract**

This study presents, for the first time, a conceptual and formal model of postsingular science (PSS), which analyses and interprets changes in scientific knowledge driven by accelerating technological progress, singularity, and the integration of artificial intelligence (AI) into scientific processes. The PSS model is based on the interplay of six key components: cumulative knowledge, intelligence, technological synergy, quantum information, social dynamics, and environmental sustainability. The interaction of these variables is described through a system of nonlinear differential equations, reflecting the complex feedback loops and synergetic effects characteristic of the postsingular world.

A differentiation table contrasting postsingular and classical science is also presented, highlighting the most fundamental differences between contemporary classical science and future postsingular science.

The model emphasizes the synergy between humans and artificial intelligence, the role of quantum technologies in accelerating scientific discovery, and the impact of social and ecological factors that either constrain or stimulate scientific progress.

It is anticipated that new forms of scientific information dissemination will replace traditional academic publications and that scientific processing will reach an entirely new level of development following the singularity-driven acceleration of technological progress and the integration of AI into R&D. This will herald an era of nonstop, ultrarapid science operating 24/7. The synergy of humans and artificial intelligence will create a scientific union on the basis of fundamentally new principles and methods.

This research provides an initial theoretical foundation for further interdisciplinary studies aimed at developing sustainable strategies and effectively managing scientific progress in the postsingular era. We believe that the actual postsingular era will occur within the next 5-7 years. Therefore, it is essential to begin developing a science policy for the postsingul science field today.

**Declarations and Statements:**
No conflicts of interest
This work was not funded
No competing or financial interests
All the data used in the work are in the public domain
Ethics committee approval is not needed (without human or animal participation).
Generative AI (LLM or other) was not used in writing the article.

---

[1] Fellow of the Royal Asiatic Society of Great Britain and Ireland

# 1. Introduction

Long ago, in the early 1990s, a major international debate on creationism took place at the pre-Putin Moscow State University (Thagard, 1989). The audience—scientists, graduate students, and undergraduates, including myself—watched with interest as evolutionists and creationists engaged in heated arguments. I don't recall the conclusion of the debate, but it ended relatively amicably. After all, truth is rarely born in disputes.

Today, if natural life and intelligence arose evolutionarily, then artificial life and intelligence represent pure creationism. Artificial intelligence (AI) emerged as an act of supernatural creation by gods in the guise of *Homo Scientificum*. The development of AI involves its own internal evolution, but natural intelligence was not created instantly either.

Until 2020, discussions about AI were cautious yet optimistic. Over the past five years, such restraint has vanished. Even though true AI does not yet exist, it is spoken of as a fait accompli. Artificial intelligence (more accurately, quasi-intelligence) has firmly entered even the realm of everyday life, and scientific research without mention of AI is now perceived as outdated and ignorant. Few actually understand what AI actually is, yet its advocates abound. The everyday integration of AI already raises concerns about its excessive role in society and, particularly, in science (Grech et al., 2023).

Postsingular science, of course, transcends traditional approaches to research and knowledge creation. The issue here is not merely a comparison between artificial and natural intelligence. It is about the scientific synergy between two levels of intellect and methodologies in scientific activity.

We argue that, in the era of postsingular science, the lines of hypothetical synergy will run through the interaction of several fundamental factors underlying any science: cumulative knowledge, collective intelligence, technological synergy, quantum information, social dynamics, and environmental sustainability.

When nonlinear feedback loops, saturation effects, and phase transitions are introduced into this set of factors, what emerges is precisely what we call postsingular science (PSS).

Traditional approaches to modelling the science of the future often assume linear methods. Despite the extensive body of work on technological singularity, the structure and system of science in this future period remain largely unexplored.

We examine the initial principles of postsingular science both from foundational conceptual positions and through an initial formalism.

The formal model of postsingular science proposed here may be useful not only for science as a whole but also for developing preventive solutions and policies in the fields of science, technology, and sustainable development.

## 2. Literature Review

The concept of "postsingular science" does not appear explicitly in the scientific literature, at least on the basis of searches in Dimensions and Mendeley. It is possible that this conceptual framework exists under other terminologies. Nevertheless, the paradigms of postsingular science align closely with the primary contexts and narratives of technological singularity, postsingular society, and the evolution of scientific artificial intelligence.

Digital technologies, powered by artificial intelligence, have become a solid foundation for scientific and technological progress (*Introduction to Digital Humanism*, 2024). The continued development of AI will inevitably lead to a scenario where humanity cedes its civilizational primacy to nonhuman intelligence (Palm, 2023).

Regarding the advancement of science, we posit that AI's primary advantage over humans in scientific activities lies in its capacity to work on scientific problems 24 hours a day. Even with current AI capabilities, this is a significant benefit, not to mention the future potential of artificial scientists. When postsingular artificial identities (Fusté, 2022) independently produce over 90% of all scientific output in the natural and social sciences, this transformation will be undeniable.

Thus, we can consider postsingularity as a systemic state that emerges after the onset of technological singularity. This concept, initially introduced by John von Neumann and further developed by Vernor Vinge (Vinge, 1993) in the context of general development issues, describes a world where machine intelligence overwhelmingly surpasses human intelligence with all the ensuing consequences.

This state is characterized by exponential growth in collective knowledge (Kurzweil, 2014), which was previously limited by the constraints of human cognition. These developments raise pressing questions about the structure of ethical principles and cultural boundaries to regulate AI in the context of scientific and technological ethics (Jinsong, 2023). This is particularly relevant to transhumanism (Al-Kassimi, 2023). However, the overarching concern remains the general safety of integrating AI into the human community (Bostrom, 2014). This challenge is amplified by the clear advantages that AI technologies have over human technologies (Makridakis, 2017), which require humanity to adapt radically to new social conditions (Bartlett, 2009).

## 3. Results

### 3.1. *The Great Scientific Differentiation*

What are the differences between contemporary classical science and postsingular science in the future?

Postsingular science (PSS) can be interpreted as a scientific system that emerged after the achievement of technological singularity—an event marking the point at which artificial intelligence and other technologies begin to evolve autonomously and exponentially. Unlike contemporary classical science, PSS is characterized by fundamental shifts in methodologies, organizational structures, and philosophical foundations of scientific inquiry.

Below, we outline the fundamental differences between PSSs and modern classical science:

*Table 1.* Differentiation of postsingular and classical sciences

| **Postsingularity science** | **Modern classical science** |
|---|---|
| *Publication activity* ||
| Instead of publications, scientific data are interpreted through epistemic tokens | Scientific results are interpreted through printed and electronic publications in IMRAD format |
| Epistemic tokens contain systemic and structured results in their pure form | The publications contain a large portion of irrelevant information that is additional to the results |
| Epistemic tokens are localized in the scientific data integrator | Publications and scientific content are localized in scientific journals and publications |
| Scientific information in the form of epistemic tokens is transferred directly to the final beneficiaries who have a direct or indirect relation to scientific results | Scientific information is disseminated in journals and publicly available publications |
| *The goals and values of science* ||
| Focus on creating self-evolving systems that go beyond human needs and are oriented towards long-term survival and evolution of the mind | Obtaining fundamental and applied knowledge. Solving practical problems |
| The erosion of the anthropocentric approach, where the system of interactions between intelligence, technology and the environment is at the center | The dominance of the anthropocentric approach, in which science serves the interests of humanity |
| *Intellectual basis* ||
| Integration of human intelligence with artificial superintelligence to create a unified cognitive system | Based on human intelligence as the main tool for analysis, interpretation and generation of new knowledge |
| Artificial intelligence as an autonomous subject of knowledge, analysing its theories and generating its own research | Scientists use artificial intelligence (AI) systems as auxiliary tools for computing, data analysis and process automation |
| The interaction between humans and machines is moving from an instrumental level to a synergistic one, where AI and humans enrich or replace each other | Scientific discoveries occur through the intellectual efforts, experiments, and theoretical work of researchers who are limited by resources and computing power |
| *Methods and approaches* | |
| Mainly quantum computing, which enables the processing of enormous amounts of data in real time and opens up new ways of studying and modelling complex systems | Scientists use analytical methods, empirical research, modelling and experiments |
| With simulations and virtual experiments to explore hypothetical scenarios and systems that are inaccessible to empirical observation | Scientific research is limited by computing power, available resources, and the temporal costs of conducting research |
| The methods are nonlinear and multidimensional, which allows taking into account complex interactions and effects that were forced to be ignored in classical approaches. | A scientist's toolkit includes standard scientific instruments, computers and specialized software |
| *The pace and scale of progress* | |
| Exponential growth rates thanks to superfast intelligence, automated processes and self-developing systems | Science develops in a linear or limitedly accelerated mode. Discoveries and inventions take time, determined by cognitive and resource limitations and capabilities. |
| New knowledge is generated continuously by autonomous research systems that are able to operate without human intervention around the clock. | Limited by human ability to process information, conduct experiments, provide social functions, and not go beyond biological capabilities |
| The scale of scientific activity of one scientific system covers all spheres and worlds for possible research | Scientific achievements are differentiated and require significant time and material costs even within one scientific field |
| *Organization of science* | |
| Decentralized by artificial intelligence networks that can autonomously coordinate their actions | Organized around academic and research institutions, corporate laboratories and government agencies |
| Distributed research systems where scientists, AI and machine agents work together in real time | Requires significant human resources, centralized management and funding |
| The economy of science is moving towards more automated models, where funding and resources are managed by algorithmic systems. Lack of publication activity. Knowledge transfer occurs according to different principles. | Hierarchical structure of research organization. Scientific societies rely on publications, conferences and grant funding to coordinate research |
| *Object of study* | |
| objects of research extend beyond traditional physical and social systems, including quantum phenomena, virtual realities and hypothetical structures | Focus on the study of nature, society, technology and physical reality, limited by three dimensions and limited by imagination |
| Holistic approaches to the study of systems as integral and interconnected entities | Uses linear and reductionist approaches, breaking down complex systems into their individual components |
| Hyperdimensional structures, alternate realities and new levels of fundamental physics are explored | Relatively simple forms of matter and entities are explored |
| *The role of data* | |
| Processing of "hyperdata" – infinitely large arrays of information generated by artificial systems in real time | Based on the collection, processing and analysis of data with limitations on the speed of researchers and the availability of technologies |
| Data scarcity is replaced by a challenge to filter, process and integrate available information. | Data shortages for many areas with significant data mining costs |
| Quantum systems and AI enable analysis previously unavailable due to computational limitations | |
| *Ethical and social aspects* | |

| New ethical issues related to the autonomy of artificial intelligence, the rights of machine systems and the impact of technology on human nature | Ethical issues such as the use of genetic engineering, artificial intelligence or nuclear technology |
|---|---|
| Society becomes part of the scientific system, where the boundaries between researchers, machines and social structures are erased | They are considered primarily within the framework of human values and legal norms. |
| Ethical frameworks must include new principles that take into account the interaction between humans, artificial intelligence and the environment | Scientific discoveries influence society, but society remains the main subject of control over science. |

Postsingular science represents a new stage in the evolution of knowledge, where classical principles of scientific activity are transformed under the influence of artificial intelligence, quantum technologies, and the accelerating pace of progress. Its key differences from classical science lie in the integration of human and artificial intelligence, scaling of research processes, changes in organizational structures, and approaches to interaction with society and nature. This shift reveals unprecedented opportunities for scientific progress but also necessitates addressing new ethical, social, and environmental challenges.

Post-Singular science differs significantly from classical modern science across multiple dimensions, including the integration of artificial intelligence, research methods, the pace and scale of progress, organizational structures, and philosophical foundations. These distinctions are driven by the achievement of technological singularity, which radically transforms scientific inquiry and its role in society. Post-Singular Science opens new horizons for research but simultaneously creates challenges related to managing complex systems, ethical considerations, and sustainable development.

All these changes have led to a reassessment of the usual epistemological, ethical, and organizational frameworks of science, presenting humanity with new challenges in terms of regulation, developmental goals, and global sustainability.

With respect to publication activities, the celebratory era of scientific publications, the IMRAD structure, accolades, and honors will eventually come to an end. Instead, the routine of pure scientific inquiry will prevail, where science will cease to be solely a domain of human endeavor. Practically all scientific achievements and discoveries will no longer belong to humanity alone.

### 3.2. *Post-Singular Formalism*

Postsingular science (PSS) models the interaction of key factors in science, technology, intelligence, quantum effects, and social dynamics, which are unified within the conditions of a postsingular world. Naturally, this requires a nonlinear model that accounts for exponential growth, feedback loops, and global synergy among components.

Let us consider the fundamental variables of postsingular science: $K(t)$ (the level of cumulative knowledge), $I(t)$ (the level of integrated intelligence), $T(t)$ (technological synergy), $Q(t)$ (quantum information), $S(t)$ (social dynamics) and $E(t)$ (ecological sustainability).

These components form the core structure around which postsingular science is organized and systematized.

We propose a relatively simple yet authentic formal model that links the dynamics of all these factors into a unified system, serving as a framework for postsingular science. To achieve this, we consider the formal components of each factor individually under the simplifying assumption that each factor evolves nonlinearly and under the influence of other factors. In this system, the factors exhibit a certain degree of interconnectivity.

### 3.3. *Cumulative knowledge*

In postsingular science, knowledge is defined as the cumulative body of scientific and technological knowledge generated through the interaction of intelligence, technology, and available informational resources, including quantum information. However, the growth of knowledge is constrained by a saturation effect, where the increasing volume of information encounters limitations in processing and assimilation.

The level of intelligence (I) directly influences the creation and assimilation of new knowledge. The greater the intelligence of the system (encompassing both humans and AI), the greater the volume of knowledge that can be produced per unit of time.

Technological synergy accelerates access to new knowledge. For example, communication technologies, analytics, and data processing enable faster accumulation of knowledge. The effect of technology is represented through the function $\log(1+T)$, which reflects the diminishing returns in knowledge growth as technological synergy increases. Once technologies reach a high level, their contribution increases more slowly (saturation effect).

We introduce the coefficient $\alpha(k)$, which expresses the efficiency of intelligence in generating knowledge given the technological context. Thus, $\alpha(k)I \cdot \log(1+T)$ describes the combined impact of intelligence and technology on knowledge growth.

Quantum computing and effects, such as superposition and entanglement, enable the processing of vast amounts of data and the extraction of insights that are inaccessible to classical systems. Therefore, Q proportionally increases K, represented as β(k)Q, where β(k) is the efficiency of converting quantum information into knowledge. This coefficient reflects the impact of quantum effects on knowledge growth.

Societal factors may limit knowledge growth due to constraints such as censorship, an unequal distribution of resources, or a lack of cooperation. Similarly, ecological limitations, such as resource scarcity or crises, can slow the development of science, technology, and the creation of new knowledge.

In this context, the constraint function 1/(1+S+E) describes the diminishing impact of socioecological factors:

$$-\gamma(k)\left(\frac{1}{1+S+E}\right)$$

Here, γ(k) represents the sensitivity of knowledge to socioecological constraints. When S+E is high, the impact becomes significantly inhibitory, and 1/(1+S+E) becomes small. However, when S+E is low, their negative impact is minimal.

By combining all three components, we derive the dynamic equation for postsingular cumulative knowledge:

$$\frac{dK}{dt} = \alpha(k)\,\mathrm{I}\cdot\log(1+T) + \beta(k)Q - \gamma(k)\frac{1}{1+S+E}$$

The equation represents the dynamics of cumulative knowledge in the postsingular science environment:
1. The first term, α(k)I·log(1+T), denotes the contribution of intelligence and technology.
2. The second term, β(k)Q, signifies the contribution of quantum information.
3. The third term,
$$-\gamma(k)(1/1+S+E)$$
, accounts for the constraints imposed by society and ecology.

This equation integrates the key factors influencing the growth of knowledge in a postsingular context and their formal analogues.

### 3.4. *Intelligence*

In reality, a large volume of knowledge does not necessarily correlate with intelligence. A person may possess exceptional mnemonic abilities and an extensive repository of knowledge, yet such erudition can also coexist with a low level of intelligence.

However, in the postsingularity context, intelligence (I) is directly proportional to the level of knowledge (K), as any volume of information is authentically adapted and interpreted. Postsingular knowledge becomes dynamic—without dynamics, knowledge cannot exist. Consequently, the more knowledge a system has, the greater its potential for intellectual activity.

The smoothing of knowledge growth is interpreted through the expression $K/(1+K)$. In this case, the function $K/(1+K)$ describes the saturation effect. Specifically, for low values of K, knowledge linearly increases intelligence $K/(1+K) \approx K$. The high value of the growth of intelligence slows down because expanding the knowledge base requires increasingly more time for processing and integration, even within postsingular science.

We introduce the coefficient $\alpha(i)$, which reflects the efficiency with which knowledge is transformed into intelligence. The expression $\alpha(i) \cdot (K/(1+K))$ thus describes the impact of knowledge on intelligence, accounting for the saturation effect.

Technologies (T) enhance intelligence by automating processes, analytics, machine learning, and other tools. Their contribution to intelligence is interpreted through the function $\sqrt{T}$.

Instead of a linear relationship, the square root reflects diminishing returns. In the early stages, technologies significantly accelerate intellectual activity, but as their saturation increases, their influence naturally diminishes.

We introduce the coefficient $\beta(i)$, which indicates the effectiveness of technology in influencing intelligence. The expression $\beta(i)\sqrt{T}$ captures the contribution of technologies to intellectual development, considering the diminishing returns effect.

Quantum technologies accelerate data processing, but their excessive use can overload intellectual systems. For example, processing massive volumes of hyperdata demands significant resources and may limit the efficiency of intelligence.

In this case, an inverse relationship of $1/(1+Q)$ is appropriate, where Q represents quantum information as the standard informational foundation of postsingular science. For low values of Q, the constraint is minimal ($1/(1+Q) \approx 1$), whereas for high values, the constraint becomes more pronounced ($1/(1+Q) \to 0$.

Thus, $-\gamma(i)(1/1+Q)$ accounts for the negative effect of excessive quantum data on intelligence. Here, the coefficient $\gamma(i)$ represents the sensitivity of intelligence to excessive intellectual and machine data processing loads.

We assume that the change in intelligence I in postsingular science will grow due to the accumulation of knowledge K and technological synergy T.

However, the dynamics of intelligence are constrained by the overload of quantum information Q.

Therefore, the dynamic equation for intelligence can be formally expressed as follows:

$$\frac{dI}{dt} = \alpha(i)\,\frac{K}{1+K} + \beta(i)T - \gamma(i)\frac{1}{1+Q}$$

The equation for intelligence I(t) models the dynamics of cumulative intelligence in a system that combines human, artificial, and synergistic aspects. Its structure incorporates key factors influencing the development of intelligence and is expressed as a sum of positive contributions (knowledge and technology) and limiting factors (overload from quantum data).

Overall, the equation reflects the basic dynamics of postsingular intelligence, characterized by exponential knowledge growth, diminishing returns from technologies, and saturation effects.

### 3.5. *Technological Synergy*

Modern technologies often exhibit exponential growth, as described by empirical laws such as Moore's Law (e.g., the growth of processing power). While Moore's law has become outdated and inaccurate today, its conceptual foundation remains relevant.

Intelligence (I) accelerates technological development. The higher the level of combined intelligence (including human and artificial intelligence) is, the faster technologies evolve. Consequently, the exponential dependency (exp(β(t)I) reflects the accelerating influence of intelligence on technological progress, in line with Kurzweil's Law of Accelerating Returns (Kurzweil, 2005). The current level of technology (T) acts as a multiplicative factor, representing the idea that more advanced technologies provide a platform for faster growth. While this concept may appear straightforward, in postsingular science, it achieves an absolute parity between qualitative development and quantitative growth.

We introduce the coefficient α(t), which defines the baseline speed of technological growth. This coefficient depends on factors such as available resources, the degree of global cooperation, and other conditions. The expression:

$$\alpha(t)T \cdot e^{\beta(t)I}$$

describes the exponential acceleration of technological growth driven by intelligence.

However, technological growth is constrained by ecological factors (E). For example, the resources required for manufacturing high-tech components (e.g., rare-earth metals and energy) may become scarce. In a postsingular society, ecological factors are expected to take precedence over technological growth factors. These ecological constraints can be interpreted through a simple function:

$$\frac{1}{1+E}$$

For low values of E (ecological stability), the constraint is minimal (1/(1+E)→1). For high values of E (ecological crises), the impact becomes significant (1/(1+E)→0).

We introduce the coefficient γ(t), which quantifies the extent to which ecological constraints impede technological progress. The expression:

$$-\gamma(t)\frac{1}{1+E}$$

accounts for the negative (or rather limiting) effect of ecological factors on technological development.

Considering exponential growth and ecological constraints, we arrive at the dynamic equation for technological synergy:

$$\frac{dT}{dt} = \alpha(t)T \cdot e^{\beta(t)I} - \gamma(t)\frac{1}{1+E}$$

The positive term represents the acceleration of technological growth under the influence of intelligence. The negative term reflects the constraints imposed by ecological factors.

Technological synergy (T) describes the combined impact of interconnected technologies and their influence on scientific and intellectual progress. The equation accounts for the exponential growth of technologies and the synergistic effects of intelligence (I) and technology (T) on each other.

At early stages, when technologies and intelligence are in their nascent development phases, the contribution of exponential growth is significant, whereas that of ecological constraints is negligible. As technologies advance, their growth slows due to ecological limitations, even if intelligence continues to grow. This phenomenon is understandable given the finite amount of ecological resources available to a population of 8 billion and the energy demands of technologies, whereas the intellectual resources in postsingular science remain virtually unlimited.

### *3.6. Quantum Information*

Technologies (T) act as catalysts for the development of quantum information, providing tools for quantum computing, quantum networks, and the management of quantum states. Intelligence (I) determines how effectively these technologies can be utilized to extract quantum information. Both human and artificial intelligence drive the development of algorithms, models, and methods for working with quantum systems. In this sense, the straightforward linear term T·I represents the synergistic effect of the interaction between technology and intelligence. If either T or I is low, the contribution to Q(t) will be limited.

By rewriting the synergy effect as $\sqrt{T \cdot I}$, we can account for the diminishing returns effect. That is, at initial stages, the synergy between technology and intelligence leads to significant growth in quantum information, but as saturation occurs, the effect slows down. This is evident and, in some sense, reasonable—constant growth is often harmful to optimal systems.

We introduce the coefficient α(q), which determines the rate at which the synergy of technology and intelligence is converted into quantum information:

$$\alpha(q)\sqrt{T \cdot I}$$

This expression describes the positive contribution of the synergy between technology and intelligence to the development of quantum information. The value of α(q) depends on the current state of quantum technology development.

The formation and scalability of quantum information predominantly depend on social dynamics (S), such as investment shortages, political barriers, and the lack of sufficient natural and artificial capacities for data processing. This influence can be modelled through the inverse relationship 1/(1+S), which represents the effect of social factors on the dynamics of quantum information.

When S is low (high cooperation, strong support for science), the constraint is minimal (1/(1+S)→1). When S is high (social issues, lack of collaboration), the constraint becomes significant (1/(1+S)→0).

We introduce the coefficient β(q), which reflects the sensitivity of quantum information to social constraints. The greater β(q) is, the stronger the social dynamics limit the growth of quantum information. Thus, the term:

$$-\beta(q)\frac{1}{1+S}$$

accounts for the inhibitory effect of social factors on the development of quantum information.

Taking the above into account, the dynamics of quantum information in postsingular science can be represented by the following equation:

$$\frac{dQ}{dt} = \alpha(q)\sqrt{T \cdot I} - \beta(q)\frac{1}{1+S}$$

Here, quantum information Q(t) describes the dynamics of using quantum technologies and processes in scientific computation and data processing. It is a variable that grows under the influence of technology and intelligence but may be constrained by social factors. The first term represents the positive contribution of the synergy between technology and intelligence to quantum information. The second term reflects the inhibitory effect of social factors on the growth of quantum information.

This equation considers that quantum information does not evolve in isolation but within a framework of social dynamics that can either accelerate or constrain progress. It accounts for the interplay between technological advancements, intelligence, and societal conditions, offering a balanced view of quantum information development in the postsingularity era.

### 3.7. *Social Dynamics*

Knowledge does not always increase social stability (S), but an informed society is generally more stable than an ignorant society. Additionally, knowledge enhances a society's ability to adapt to new environments and challenges.

We introduce the coefficient $\alpha(s)$, which represents the strength of the influence of cumulative knowledge on the social dynamics of a postsingular society. Its value depends on the level of integration of knowledge within society. The positive effect of knowledge (K) on social dynamics, considering the diminishing returns effect, can be expressed as:

$$\alpha(s) \cdot log(1 + K)$$

Ecological stability contributes to social harmony, reduces competition for resources, and improves quality of life. We introduce the coefficient $\beta(s)$, which reflects the sensitivity of a postsingular social system to ecological conditions. The proportional linear relationship between ecological conditions (E) and social dynamics can be expressed as $\beta(s) \cdot E$

An excessively high level of intelligence (I) can create imbalances in the social system. For example, overautomation or the concentration of intellectual resources can exacerbate inequality, reduce public engagement in societal processes, and challenge existing social institutions through the rapid development of AI. Therefore, we consider a limiting function of the form I/1+I

The degree of negative impact of concentrated intelligence on social dynamics can be interpreted as $-\gamma(s)(I/1+I)$, where $\gamma(s)$ is a coefficient that quantifies the influence of intelligence on social dynamics. Societies with higher levels of cooperation are less sensitive to the concentration of intelligence.

Thus, the equation for social dynamics is as follows:

$$\frac{dS}{dt} = \alpha(s)\log(1+K) + \beta(s)E - \gamma(s)\frac{I}{1+I}$$

Social dynamics S(t) reflect the influence of societal factors, including cooperation, competition, politics, and economics, on the development of science, technology, and knowledge. The equation for S(t) incorporates three key contributions: the positive influence of knowledge (K), support from ecological stability (E) and the limiting influence of concentrated intelligence (I) can disrupt social balance.

Social dynamics are shaped by both stimulating (knowledge and ecology) and limiting (intelligence concentration) factors. They are sensitive to the state of knowledge, ecological sustainability, and distribution of intelligence within society. This equation captures the balance of forces that influence social cohesion and adaptability in a postsingular environment.

### 3.8. Ecology

Social dynamics (S), encompassing societal awareness, environmental programs, and investments in sustainable development, contribute to improving the state of ecology (E). A high level of S can encourage the adoption of environmental standards, the implementation of clean technologies, and the use of renewable resources. The positive influence of social dynamics (S) on ecology (E) is linear and expressed as:

$$\alpha(e) \cdot S$$

Here, α(e) represents the efficiency of the social system in improving the environmental situation, which depends on the level of environmental education, policy, and international cooperation.

Knowledge (K) generally promotes more rational resource use, provided that it is not constrained by overly conservative or inefficient systems. However, exponential knowledge growth can increase environmental pressures due to the energy-intensive nature of knowledge mining, production expansion, resource consumption, and the acceleration of technological cycles. For example, large language models and other AI systems consume significant resources and often lack ergonomic efficiency.

We model the limiting effect of knowledge on ecology via the function K/(1+K), which captures the saturation effect: for low K, the impact on ecology increases linearly (K/(1+K)≈K), whereas at high KKK, the growth effect decreases (K/(1+K)→1).

The negative impact of cumulative knowledge as a major energy and resource consumer on environmental sustainability is represented as −β(e)(K/1+K). where β(e) indicates how strongly knowledge increases ecological pressure. Its value depends on the balance between resource efficiency and resource depletion or limitation.

Technologies (T) accelerate the use of natural resources (e.g., mining, energy production) and increase pollution levels (e.g., greenhouse gas emissions, plastic waste discharge into ecosystems). We model technological pressure via the function T/(1+T). For low T, the impact increases significantly (T/(1+T)≈T), whereas at high T, the effect decreases (T/(1+T)→1).

The negative influence of technological progress on environmental sustainability is represented as follows:

$$-\gamma(e)\left(\frac{T}{1+T}\right)$$

where γ(e) quantifies the impact of technologies on ecology, with its value depending on the nature of the technologies. For energy-efficient or environmentally friendly technologies, the impact will be smaller.

Summing these three components, the dynamic equation for postsingular ecology is as follows:

$$\frac{dE}{dt} = \alpha(e)S - \beta(e)\frac{K}{1+K} - \gamma(e)\frac{T}{1+T}$$

Ecology (E(t)) models the state of the environment and its sustainability. This state is determined by a balance of positive factors (social efforts, sustainable development) and negative factors (resource depletion, pollution, and technological pressure). The equation for E includes three key components: the positive influence of social dynamics (S). the negative influence of knowledge growth (K) is tempered by a saturation effect and the negative influence of technology growth (T) is also subject to diminishing returns.

### 3.9. *Final System of Equations*

The complete system of postsingular science in its simplified form is described by a set of nonlinear differential equations.

Knowledge K grows through the synergy of intelligence (I), technologies (T), and quantum information (Q), with diminishing returns from accumulated knowledge:

$$\frac{dK}{dt} = \alpha(k)\,I \cdot \log(1 + T) + \beta(k)Q - \gamma(k)\frac{1}{1 + S + E}$$

Intelligence I is driven by the growth of knowledge (K) and technologies (T) but limited by diminishing returns:

$$\frac{dI}{dt} = \alpha(i)\,\frac{K}{1 + K} + \beta(i)T - \gamma(i)\frac{1}{1 + Q}$$

Technologies T are enhanced by quantum information (Q) and social dynamics (S), but their growth slows due to saturation:

$$\frac{dT}{dt} = \alpha(t)T \cdot e^{\beta(t)I} - \gamma(t)\frac{1}{1 + E}$$

Quantum information Q is driven by technologies (T) and their interaction with intelligence (I) but is subject to saturation effects:

$$\frac{dQ}{dt} = \alpha(q)\sqrt{T \cdot I} - \beta(q)\frac{1}{1 + S}$$

Social dynamics S depend on knowledge (K) and ecological sustainability (E), with growth slowing at high levels of intelligence (I):

$$\frac{dS}{dt} = \alpha(s)\log(1 + K) + \beta(s)E - \gamma(s)\frac{I}{1 + I}$$

Ecological sustainability E is supported by social dynamics (S) but undermined by the pressures of knowledge (K) and technologies (T):

$$\frac{dE}{dt} = \alpha(e)S - \beta(e)\frac{K}{1 + K} - \gamma(e)\frac{T}{1 + T}$$

This model provides a nonlinear and integrative representation of postsingular science, enabling the study of its dynamics and identifying optimal development paths. The system describes how the interaction of knowledge, technologies, and intelligence drives exponential growth under socioecological pressures. Nonlinear dependencies include both positive (growth-enhancing) and negative (growth-limiting) feedback loops. All the components are subject to diminishing returns, preventing infinite growth and ensuring system stability.

1. **Discussion**

We interpret postsingular science (PSS) through the hypothesis that after technological singularity—when artificial intelligence and technologies consistently surpass human capabilities—the fundamental principles of scientific inquiry are redefined and substantially modernized. Post-Singular Science operates within a symbiotic reality where humans, technology, and artificial intelligence collaborate as coauthors of knowledge.

The system of equations representing postsingular science, while appearing straightforward, illustrates that all the components within the PSS are tightly interconnected, demonstrating strong direct linkages. Unlike classical science, where such connections are less apparent, the absolute rationality inherent in postsingularity ensures that these links are maximally constructive.

In this context, intelligence distributed among humans, AI, and machines is considered a unified cognitive structure. Every cognitive system—whether human, artificial, or biological networks—will be recognized as an equal participant in the scientific process. Empirical knowledge becomes inseparable from metaphysics as observers and the observed individuals mutually permeate each other.

Post-Singular Science thus presupposes that truth is inherently multidimensional and not exclusively anthropocentric. The truth depends not only on the context of observation but also on the cognitive structure generating the hypothesis. In this framework, the laws of nature vary on the basis of the cognitive architecture of the observer. Each form of consciousness—biological, artificial, or hybrid—will produce its own interpretation of reality.

Cumulative knowledge is viewed as a dynamic structure that evolves through interactions between agents of various levels. This suggests that scientific principles will resemble fractals, where local truths reflect global truths without being identical to them.

The PSS model underscores the need for an interdisciplinary approach that integrates knowledge from neuroscience, quantum physics, artificial intelligence, social sciences, and ecology. This approach ensures a comprehensive and accurate description of the dynamics of scientific progress in singularity and postsingularity conditions.

This model necessitates the integration of six key variables, each of which play a critical role in describing scientific progress:
1. Cumulative knowledge forms the foundation of progress, which depends on intelligence and technology but is constrained by social and ecological factors. The diminishing returns effect—where knowledge growth slows at higher levels—will persist, similar to the deceleration observed in modern science over the past two decades. However, the scale of the PSS will be unprecedented.

2. Intelligence will be stimulated by new levels of knowledge and technology, but, as in classical science, it will face limitations from quantum information owing to potential data overload.
3. Quantum technologies will unlock new domains of discovery, including areas presently unimaginable, with significant breakthroughs likely in the social sphere.
4. Social dynamics are driven by knowledge and ecological sustainability but may face limitations from high concentrations of intelligence, which can cause societal imbalances due to automation or the dominance of artificial intelligence.
5. Feedback loops play a fundamental role in PSS. For example, intelligence growth may accelerate technological development, whereas rapid technological progress could lead to ecological and social constraints, slowing further growth.
6. Ecology remains a critical factor in balancing progress and sustainability.

The PSS model allows for the exploration of various scenarios for scientific and technological development under different initial conditions and external factors.

Accurate determination of the coefficients in this model requires a significant body of empirical data. Furthermore, the model does not account for stochastic processes and random fluctuations, which could significantly impact dynamics, particularly during global crises.

## 2. Conclusion

It is evident that the scientific process will reach an entirely new level of development following the singularity-driven acceleration of technological progress and the integration of artificial intelligence into R&D. This will mark the advent of ultrafast science (24/7), nonstop innovation. The synergy of humans and artificial intelligence will establish a scientific alliance operating on entirely new fundamental principles and methods.

At the core of this creative scientific alliance will be nonlinear correlations between the key variables of scientific progress and societal functions—cumulative knowledge, intelligence, technological synergy, quantum information, social dynamics, and ecological sustainability—each with its own strengths, risks, and challenges.

The ultrarapid growth of knowledge and technology will present new challenges, including potential systemic overload, deterioration of ecological sustainability, and increasing social inequality. Ethical considerations will take center stage, addressing questions of autonomy in artificial systems, their societal impacts, and global accountability for the consequences of scientific and technological progress. Interaction between humans and intelligent systems will necessitate the development of cogovernance principles, not only in science but also across postsingular society.

We believe that the actual postsingular era will occur within the next 5-7 years. Therefore, it is essential to begin developing a science policy for the postsingul science field today.